\documentclass[doublecol]{epl2} 

\usepackage{color}
\graphicspath{{.},{./Fig/}} 
\newcounter{countuwe}

\newcounter{countfil}

\title{Depinning of three-dimensional drops from wettability defects}
\shorttitle{Depinning of 3d Drops} 

\author{Ph. Beltrame\inst{1} \and P. H\"anggi\inst{1} \and U. Thiele\inst{2}}
\shortauthor{Ph. Beltrame \etal}

\institute{                    
  \inst{1} Institut f\"ur Physik, Universit\"at Augsburg, D-86135 Augsburg, Germany\\
  \inst{2} Department of Mathematical Sciences, Loughborough University, Loughborough, Leicestershire, LE11 3TU, UK
}


\abstract{Substrate defects crucially influence the onset of
  sliding drop motion under lateral driving. A finite force is
  necessary to overcome the pinning influence even of microscale
  heterogeneities.  The depinning dynamics of three-dimensional drops
  is studied for hydrophilic and hydrophobic wettability defects 
  using a long-wave evolution equation for the film thickness
  profile. It is found that the nature of the depinning transition
  explains the experimentally observed stick-slip motion.
}

\begin{document}

\maketitle
%

\section{Introduction}
\label{sec:1}
Drops sliding along a solid substrate under the influence of a lateral
force are a very common physical phenomenon. The force might be
gravity for drops on an inclined or vertical wall, centrifugal forces
for drops on a rotating disk or external shear for drops in an ambient
flow.  Note that lateral gradients in wettability, temperature or
electrical fields can as well drive sliding motion.
For smooth homogeneous substrates an arbitrarily small driving force
results in drops that move with constant velocity and shape
\cite{PFL01,Thie01,Snoe07}. Larger driving forces may lead to shape
instabilities, e.g., trailing cusps may evolve that periodically emit
small satellite drops \cite{PFL01,Thie02}.

Real substrates, however, are normally not smooth. They are rough or
have local chemical or topographical defects. Even microscopic defects
can have a strong influence on the drop dynamics. The heterogeneities
may cause stick-slip motion \cite{ScWo98,Tetal06} or roughening
\cite{GoRa01,RoJo87} of moving contact lines, and are thought to be
responsible for contact angle hysteresis \cite{deGe85,LeJo92,QAD98,Spel06}.
Note that a local variation of the driving force (e.g., electrostatic
field or temperature gradient) may play the same role as a substrate
defect.

\begin{figure}
\centering
\includegraphics[width=0.9\hsize]{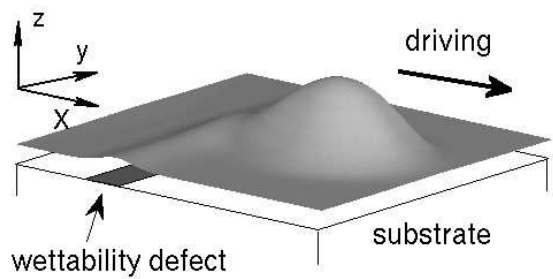}
\caption{ Sketch of the three-dimensional geometry of the problem: a
  drop on a heterogeneous substrate and under a driving force $\mu$
  along the x-direction. Thereby the heterogeneous wettability is
  assumed to depend on the x-direction spatial direction only.}
\label{fig:sketch}
\end{figure}

The present paper focuses on the depinning of three-dimensional drops
from hydrophobic and hydrophilic line defects that pin them at their
front and back, respectively: A hydrophobic defect is less wettable
for the drop that therefore has to be forced to pass it. On the
contrary, a hydrophilic defect is more wettable and the drop has to be
forced to leave it as sketched in Fig.~\ref{fig:sketch}.  A recent
theoretical study of the depinning dynamics of less realistic
two-dimensional drops employs lubrication approximation and finds
stick-slip motion beyond depinning \cite{ThKn06,ThKn06b}.

The present work is based on a thin film evolution equation in
long-wave approximation \cite{ODB97,KaTh07} that incorporates
wettability in the form of an additional pressure term -- the
so-called disjoining pressure \cite{deGe85}. It models the effective
molecular interactions between the substrate and the free surface of
the liquid, e.g., long-range apolar van der Waals interactions and
short-range polar electrostatic or entropic interactions
\cite{Shar93}. With the proper choice of terms such a disjoining
pressure describes the behaviour of drops of partially wetting
liquid with a small equilibrium contact angle that coexist with an
ultra-thin precursor film. An advantage of such a
model is the absence of a contact line singularity. Note that
although only small contact angles and small driving forces are
compatible with the long-wave approximation results
are often qualitatively correct for more general conditions.
Incorporating wettability in the form of a disjoining pressure allows
to study the influence of chemical substrate heterogeneities or
defects by a spatial modulation of the involved material parameters.
For dewetting thin films without lateral driving this is done in
\cite{KKS00,TBBB03}.

The analysis of the two-dimensional problem in
Refs.~\cite{ThKn06,ThKn06b} consists of a study of (i) steady drops
and their stability based on continuation techniques for ordinary
differential equations \cite{DKK91} and (ii) time-periodic solutions
sliding over a regular array of defects based on 'standard'
time-stepping schemes. The here presented study of the
three-dimensional case is based on recently developed effective
algorithms for both, the continuation of pinned steady drops described
by a partial differential equation and the time simulation of the
dynamics beyond depinning \cite{BeTh08}.
\section{Model and numerical method}
\label{sec:mod}
We consider a liquid layer or drop on an inhomogeneous two-dimensional
solid substrate as sketched in Fig.~\ref{fig:sketch}. The liquid
partially wets the substrate (with a small equilibrium contact angle)
and is subject to a small constant lateral force $\mu$ that acts in
the $x-$direction. Employing the long-wave or lubrication
approximation the dimensionless evolution equation for the film
thickness profile $h(x,y,t)$ is derived from the Navier-Stokes
equations, continuity and boundary conditions (no-slip at substrate,
force equilibria at free surface) \cite{ODB97,KaTh07,ThKn06}. It reads
\begin{equation} 
  \partial_{t}h=-\nabla\cdot\left\{ m(h)\left[ 
\nabla\left(\Delta h+\Pi(h,x)\right)+\mu\mathbf{e_{x}} 
\right]\right\},\label{eq:lub}
\end{equation} 
where $\nabla=(\partial_{x},\partial_{y})$ and
$\Delta=\partial_{xx}^{2}+\partial_{yy}^{2}$ are the planar gradient
and Laplace operator, respectively. The mobility function $m(h)=h^{3}$
corresponds to Poiseuille flow and $\Delta h$ represents the Laplace
pressure (capillarity). The disjoining pressure $\Pi(h,x)$ models the
position-dependent wetting properties that in the case of transverse
line defects only depend on the streamwise direction.  The literature
discusses a plethora of different functional forms for $\Pi(h)$
\cite{deGe85,TDS88}. Most model the presence of an
ultra-thin precursor film of about 1-10 nm thickness and thereby avoid
a 'true' film rupture. We employ long-range apolar van der Waals
interactions combined with a short-range polar interaction
\cite{deGe85,Shar93,TVN01}
\begin{equation} 
  \Pi(h,x)=\frac{b}{h^{3}}-\left(1+\epsilon\xi(x)\right)e^{-h},\label{eq:dispin}
 \end{equation} 
 where $\epsilon$ and $\xi(x)$ are the amplitude and profile of the
 heterogeneity, respectively. To model a localized defect $\xi(x)$ is
 based on Jacobi elliptic functions as described in
 \cite{ThKn06,ThKn06b}. Typical examples can be seen below in
 Fig.~\ref{fig:pro}.  The amplitude $\epsilon$ represents the wettability
 contrast. For $\epsilon < 0$ [$\epsilon> 0$] the defect is less
 [more] wettable than the surrounding substrate, i.e., the defect is
 hydrophobic [hydrophilic]. Based on the Jacobi functions we study
 drops on a periodic array of defects.  The period $L_x$ is chosen
 sufficiently large to avoid interactions between subsequent
 drops/defects.  The imposed spatial periodicity allows to
 characterize stick-slip motion by its period in time.

 Based on Eq.~(\ref{eq:lub}) with (\ref{eq:dispin}) the depinning
 behaviour in the three-dimensional (3d) case is analysed following
 the methodology used in \cite{ThKn06,ThKn06b} for the two-dimensional
 (2d) case. Steady-state solutions (pinned drops) and their stability
 are determined using continuation techniques and the stick-slip
 motion beyond the depinning threshold is analysed using time-stepping
 algorithms. In the 2d case an explicit scheme suffices for the
 latter. The continuation can be performed using the package AUTO
 \cite{AUTO00} as the underlying equation corresponds to a system of
 ODE's \cite{KaTh07}.  In the 3d case an effective and exact time
 simulation of Eq.~(\ref{eq:lub}) is challenging and leads to a number
 of numerical problems \cite{BGW01,Thie02,BeGr05}.  Here we
 employ a recently developed approach \cite{BeTh08} based
 on exponential propagation \cite{FTDR89}. It allows for a very good
 estimate of the optimal timestep for the different regimes of the
 dynamics. This is of paramount importance as close to the depinning
 transition it needs to be varied over many orders of magnitude. The
 second advantage lies in the possibility to adapt the time-stepping
 scheme in a way that it can be used to continue the branches of
 steady drop states and to determine their stability. For details see
 Ref.~\cite{BeTh08}.

\section{Depinning of 2d drops}%
\label{sec:2d}
\begin{figure}
\centering
\includegraphics[width=0.45\hsize]{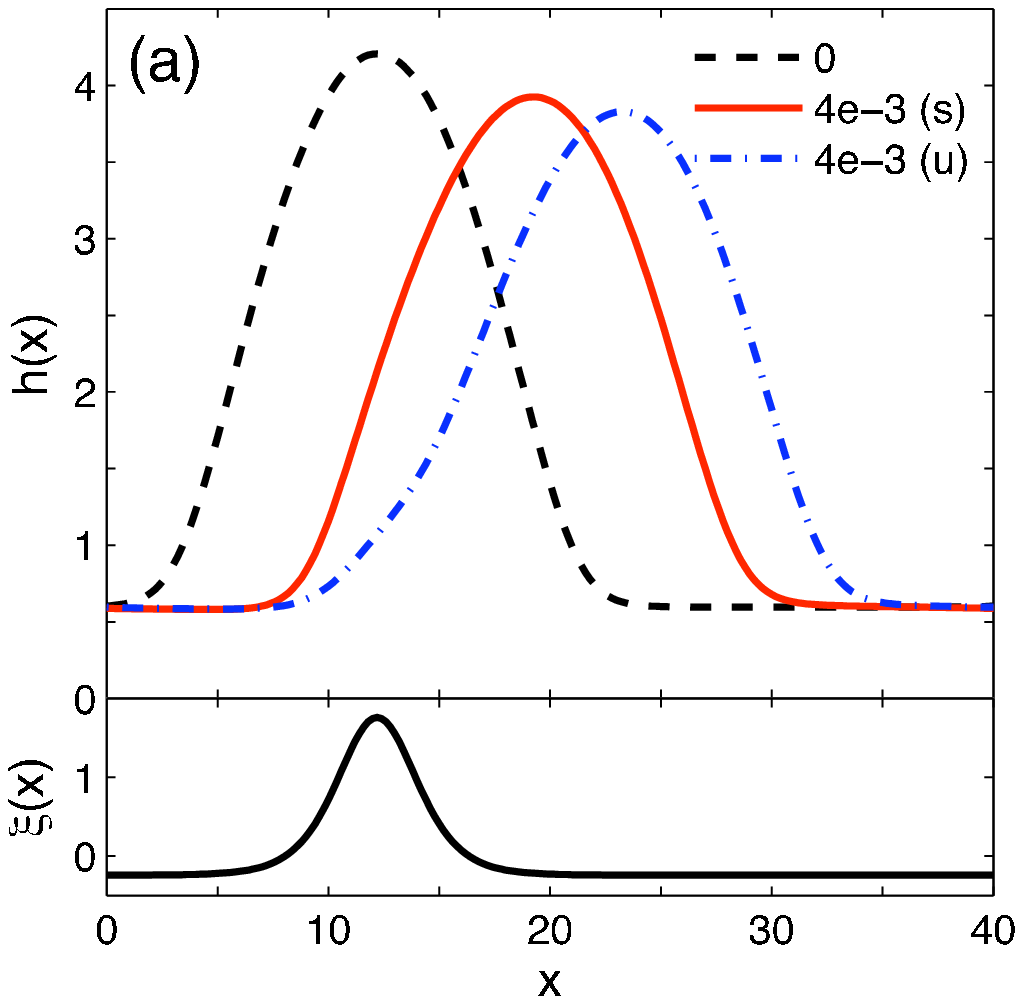} \hspace{0.05\hsize} \includegraphics[width=0.45\hsize]{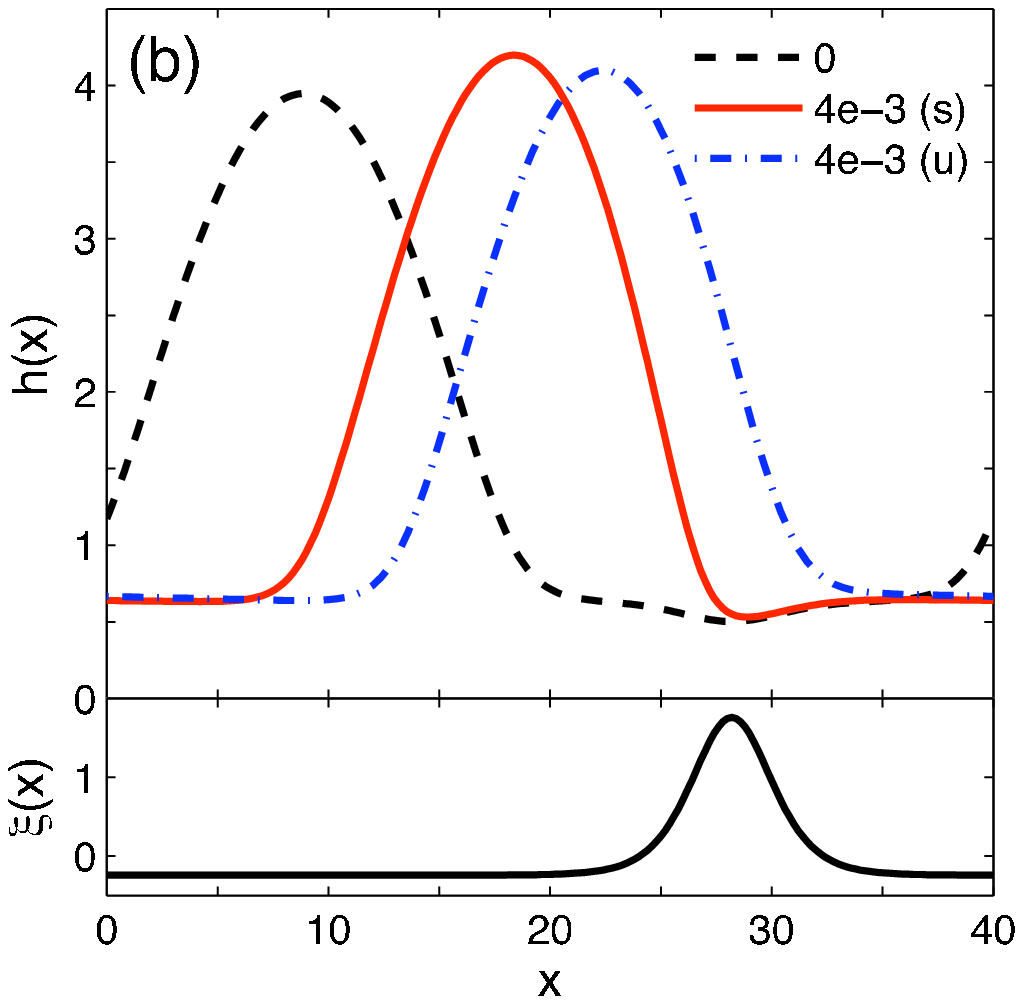}\\
\includegraphics[width=0.45\hsize]{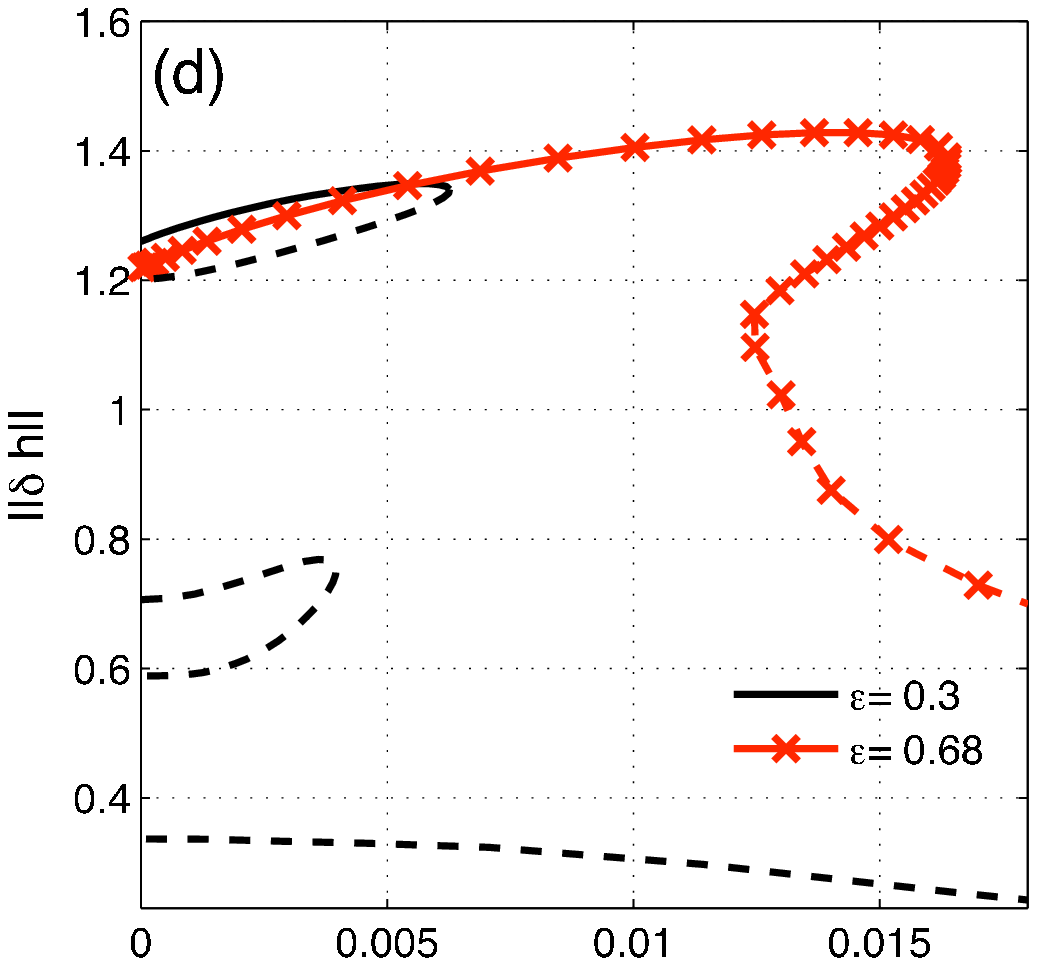}\hspace{0.05\hsize}\includegraphics[width=0.45\hsize]{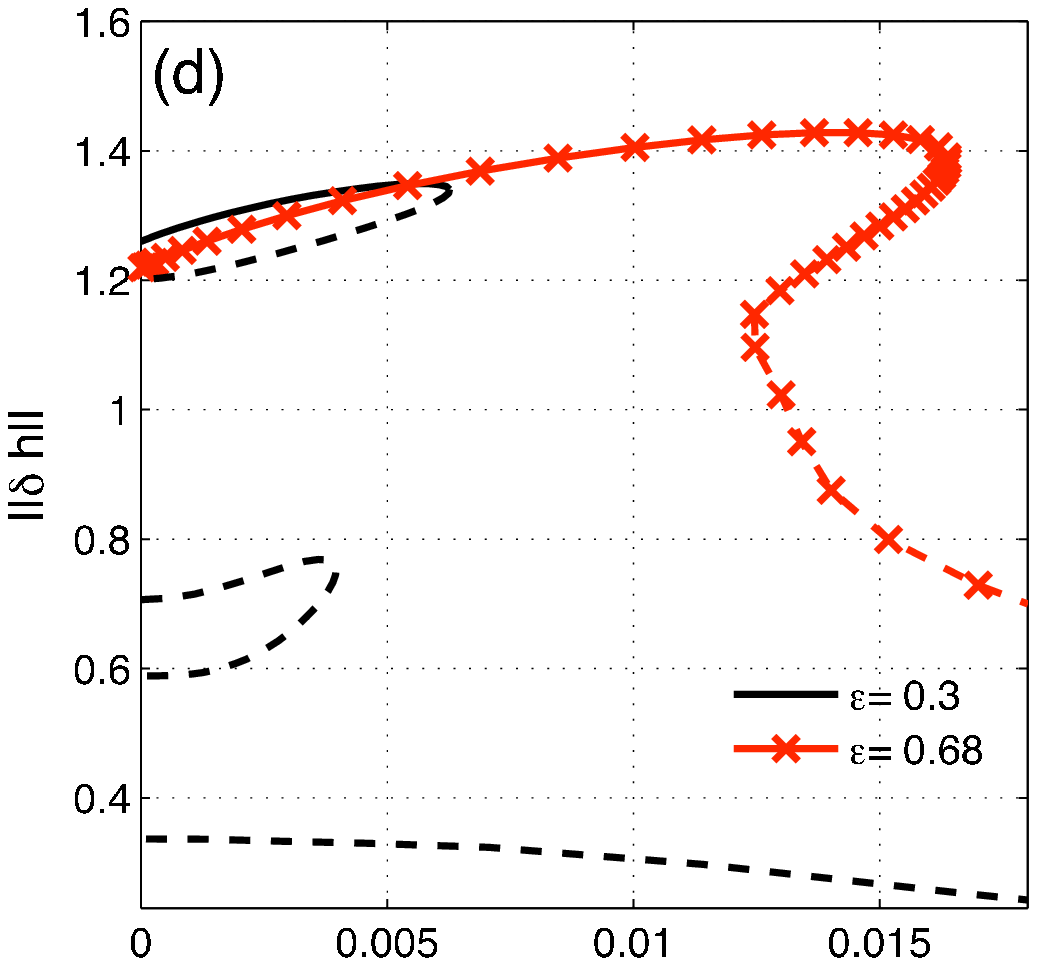}
%
\caption{Selected drop profiles (top row) and corresponding
  bifurcation diagrams (bottom row) for localized hydrophilic
  ($\epsilon = 0.3$, left column) and hydrophobic ($\epsilon = -0.3$,
  right column) defect in the 2d case.  (a) and (b) give steady
  drop profiles for several driving forces $\mu\ge0$ as given in
  the legend. For $\mu=4\cdot 10^{-3}$ stable (solid line with symbol
  ``s'') and unstable (dotted line with symbol ``u'') steady drops are
  represented. The lower part of the panels gives the heterogeneity
  profile $\xi(x)$. (c) and (d) characterize branches of steady drop
  solutions by the dependence of their $L^2$ norm ($||\delta
  h||=\sqrt{\int_0^L (h-\bar{h})^2dx/L}$) on the lateral driving force
  $\mu$ for various defect strength $\epsilon$ as given in the legend.
  Dashed lines indicate unstable solutions.  Domain length, volume and
  resulting drop height on a homogeneous substrate are $L_x =40$,
  $V=66$ and $h_{max} = 4.0$, respectively.}\label{fig:pro}
\end{figure}

Before focusing on the 3d case we shortly present results for 2d using
equivalent parameter values to allow for a qualitative and quantitative
comparison.
Without lateral force ($\mu=0$) there exists a unique stable drop
solution for each wettability contrast $\epsilon$.  The drop sits on
top of a hydrophilic defect (dashed line in Fig.~\ref{fig:pro}(a)) or
in the middle between hydrophobic defects (dashed line in
Fig.~\ref{fig:pro}(b)).  Note that other steady solutions may exist
that are normally unstable.  For an in-depth study of solutions on a
horizontal substrate (for another $\Pi(h)$) see \cite{TBBB03}.
 
Increasing the lateral driving force $\mu$ from zero the drop does not
start to slide as for the homogeneous substrate, but remains pinned by
the defect. A hydrophobic defect blocks the drop at the front, it
becomes compressed and heightens (see Fig.~\ref{fig:pro}(b)) until it
finally depins. This can best be seen in the bifurcation diagram
Fig.~\ref{fig:pro}(d) where the norm of stable and unstable steady
solutions is shown in dependence of $\mu$ for several wettability
contrasts.  The norm of the stable drop solution first increases, then
decreases slightly and the branch annihilates with an unstable one at
$\mu_c$. In contrast, a hydrophilic defect holds a drop at its back,
with increasing driving it becomes stretched and lower (see
Fig.~\ref{fig:pro}(a)) until it finally depins. The accompanying
bifurcation diagram (Fig.~\ref{fig:pro}(c)) shows that the norm
decreases till the branch annihilates with an unstable one, e.g., for 
$\epsilon = -0.3$ at $\mu_c\approx0.005$.

Beyond the critical value $\mu_c$, there exists in both cases only a
single branch of steady solutions that are all linearly unstable.  Its
norm approaches zero with increasing $\mu$ (not shown) indicating that
it corresponds to slightly modulated film solutions. This state being
unstable, a time-dependent state is expected that corresponds to
sliding drops. In the 2d case such solutions where discussed in
Ref.~\cite{ThKn06}. Related solutions and the character
of the depinning transition for the 3d case will be discussed next.
%
\section{Depinning of 3d drops}
\label{sec:3d}
\begin{figure}[tbh]
\centering
\includegraphics[width=0.8\hsize]{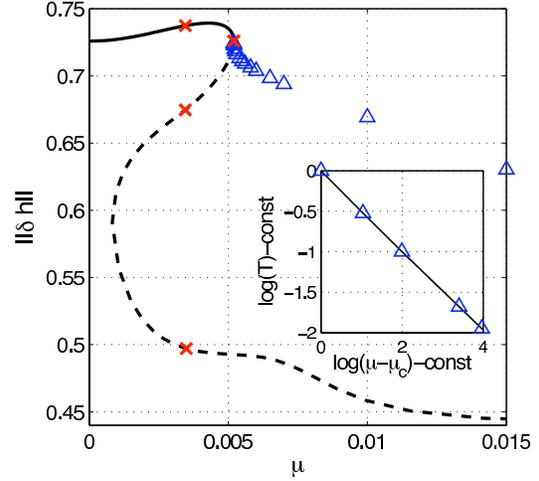}
\caption%
{Bifurcation diagram for drops pinned by a hydrophilic line defect
  of strength $\epsilon=-0.3$. Shown is the $L^2$ norm $||\delta h||$
  of steady solutions in dependence of the lateral driving force
  $\mu$.  The branch of stable pinned drops corresponds to the
  solid line whereas unstable solutions are given as dotted
  lines. Beyond the depinning bifurcation, triangles represent the
  time-averaged $L^2$ norm of time-periodic solutions that correspond
  to sliding drops performing a stick-slip motion.  The domain size
  is $40\times 40$. Crosses indicate profiles given in
  Fig.~\ref{fig:statil}. The inset gives for the stick-slip
  motion the dependence of the time-period on $\mu-\mu_c$. The
  straight line corresponds to a power law with exponent $-1/2$.  }%
\label{fig:condil}
\end{figure}

We consider now the full 3d geometry as sketched in
Fig.~\ref{fig:sketch}.  In particular, we look at hydrophilic and
hydrophobic line defects that lie orthogonal to the direction of the
driving force. In the present 3d setting one can re-interpret the
findings for 2d drops as referring to the depinning of a liquid ridge
from a line defect assuming that the transverse translational
symmetry is not broken in the depinning process.

To compare the depinning of such a ridge and the one of a true 3d drop
we use the continuation and time-stepping techniques outlined
above. Furthermore all parameters with the exception of the drop
volume are chosen as in the 2d case.  For the latter we use a value
such that the maximal drop height on a homogeneous substrate without
driving force ($\mu=0$) is equal to the one of the ridge.

\begin{figure}[tbh]
\centering 
\includegraphics[width=0.9\hsize]{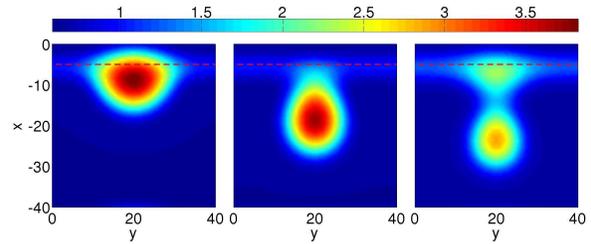}
\caption{Shown are contours of steady drop solutions for a
  hydrophilic defect for $\mu = 3.5 \cdot 10^{-3}$. Profiles from left
  to right correspond to crosses in Fig.~\ref{fig:condil} from high to
  low norm. The left panel presents the stable pinned drop.  The
  thin horizontal line marks the wettability maximum. The remaining
  parameters are as in Fig.~\ref{fig:condil}.}
\label{fig:statil} 
\end{figure}
\begin{figure}[tbh]
\centering 
\includegraphics[width=0.9\hsize]{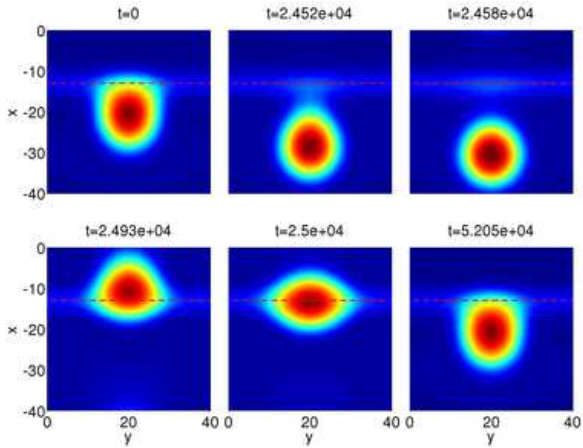}
\caption{%
  Shown are snapshots of drop profiles at different stages of a
  stick-slip cycle (at times given below the individual panels) for a
  drop depinning from a hydrophilic line defect (marked by the
  horizontal line). The chosen driving $\mu = 5.193 \cdot 10^{- 3}$ is
  still close to the critical $\mu_c$.  Color code and remaining
  parameters are as in \ref{fig:condil}. }%
\label{fig:snapsnipil} 
\end{figure}

\begin{figure}[tbh]
\centering
\includegraphics[width=0.8\hsize]{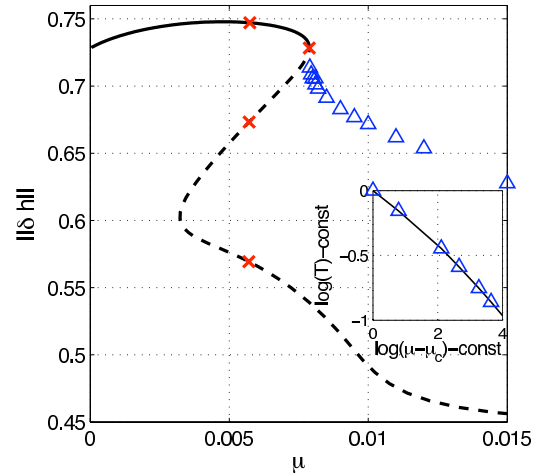}
\caption%
{Bifurcation diagram for drops pinned by a hydrophobic line defect
  of strength $\epsilon=0.3$. The presented norms, line styles,
  symbols, domain size and inset are as in
  Fig.~\ref{fig:condil}. Corresponding profiles are given in
  Fig.~\ref{fig:statob}.  The straight line in the inset corresponds
  to a power law with exponent $-1/4$.  }%
\label{fig:condob}
\end{figure}
\begin{figure}[tbh]
\centering 
\includegraphics[width=0.9\hsize]{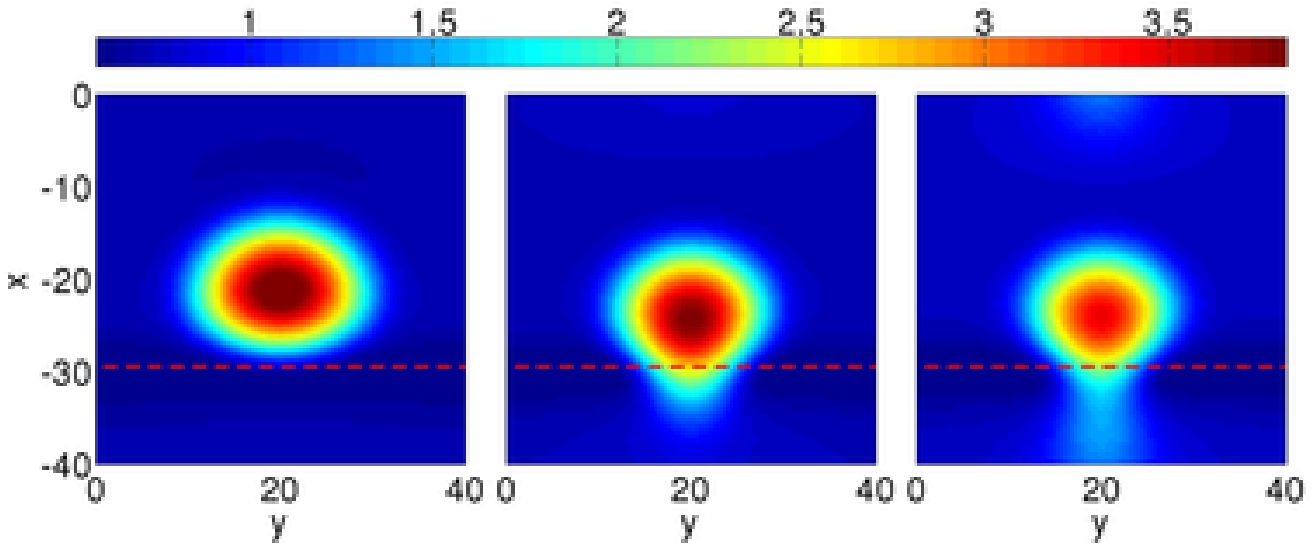}
\caption{Shown are contours of steady drop solutions for a
  hydrophobic defect for $\mu = 5.7 \cdot 10^{-3}$. Profiles from left
  to right correspond to crosses in Fig.~\ref{fig:condob} from high to
  low norm. The left panel presents the stable pinned drop.  The
  horizontal line marks the wettability minimum. The remaining
  parameters are as in Fig.~\ref{fig:condob}.}
\label{fig:statob} 
\end{figure}
\begin{figure}[tbh]
\centering 
\includegraphics[width=0.9\hsize]{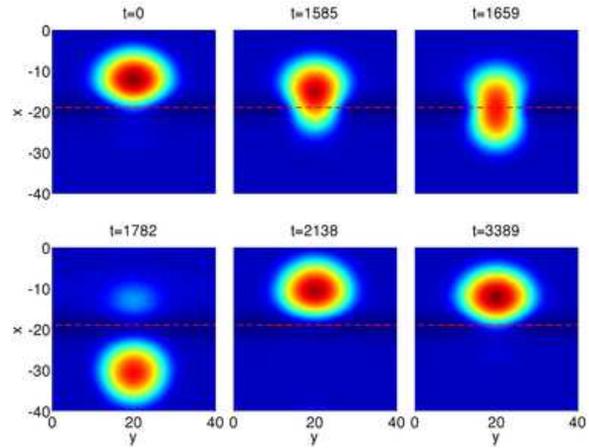}

\caption{%
  Shown are snapshots of drop profiles at different stages of a
  stick-slip cycle for a depinned drop near the depinning bifurcation
  (at $\mu = 7.898 \cdot 10^{- 3}$ and times as given below the
  individual panels) for a hydrophobic line defect (marked by the
  horizontal line).  Color code and remaining parameters are as in
  \ref{fig:condob}.
}%
\label{fig:snapsnipob} 
\end{figure}
\begin{figure}[tbh]
\centering 
\includegraphics[width=0.9\hsize]{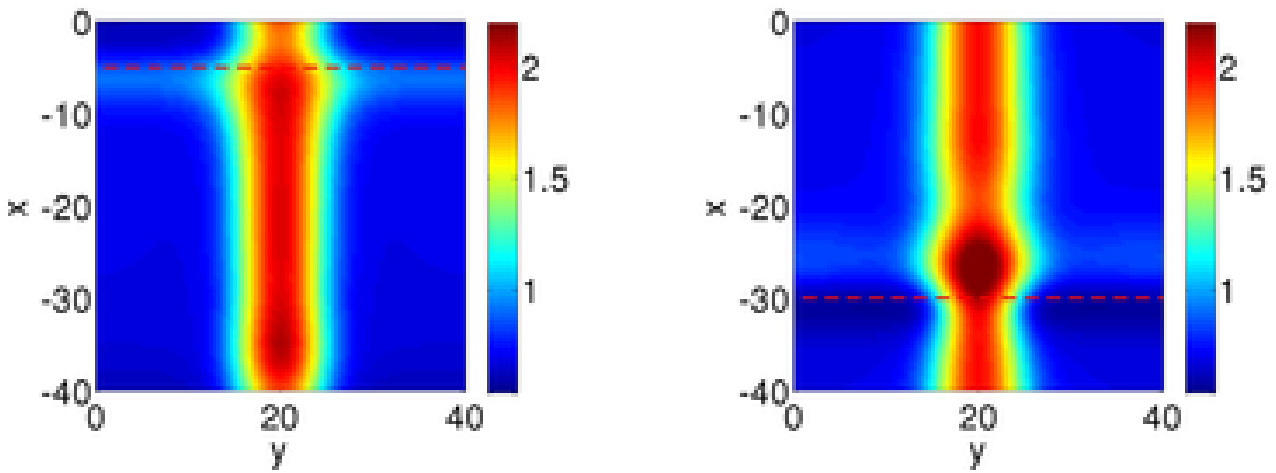}
\caption{%
  Shown are contours of steady rivulet solutions for large driving
  force $\mu=0.05$ for (a) hydrophilic and (b) hydrophobic line
  defects. The remaining parameters are as in Figs.~\ref{fig:condil},
  and \ref{fig:condob}, respectively. The horizontal line marks the
  extrema of the wettability profile.}%
\label{fig:rivulet} 
\end{figure}

Fig.~\ref{fig:condil} shows the bifurcation diagram for a single drop
on a square domain. The stable drop is pinned at its back by the
hydrophilic line defect with $\epsilon=-0.3$. On the horizontal
substrate ($\mu=0$) the drop sits symmetrically on the defect its
contour being an ellipse with the long axis on the defect (not shown).
When increasing $\mu$ the drop moves to the downstream side of the
defect where it is retained by the high wettability patch below its
tail.  With increasing $\mu$ it stretches downstream but is compressed
transversally. The combined effect of the two processes leads contrary
to the 2d case to an increase of the norm.  The stable branch loses
its stability via a saddle-node bifurcation at the critical driving
$\mu_c=5.193\cdot 10^{-3}$. The continuing unstable branch is turned
towards smaller $\mu$. It turns back again at a further saddle-node
bifurcation and the resulting 'low-norm' branch then continues towards
large $\mu$.  A selection of steady stable and unstable drop solutions
corresponding to the crosses in Fig.~\ref{fig:condil} is given in
Fig.~\ref{fig:statil}.  The left panel corresponds to the pinned
stable drop described above, the middle panel represents an unstable
drop that one could call ``at depinning'': it has an oval front shape
and is connected to the hydrophilic patch by a thin bridge that almost
looks cusp-like and seems to be at the point of breaking. Physically
it corresponds to a threshold solution: If it is moved a bit upstream
[downstream] it retracts [slips to the next defect] and converges to
the stable drop solution.  The left panel, finally, gives the unstable
solution of lowest norm. It resembles two drops joined by a thin
thread with the smaller one sitting on the heterogeneity. The
character of solutions on this branch at large $\mu$ is discussed
below.

For $\mu>\mu_c$ no steady stable solutions exist and we expect the
system to exhibit a time-dependent behaviour. In particular, we expect
in the present spatially periodic setting that drops depin from one
hydrophilic defect and slide to the next one. There they do not stop
but only slow down as the defect tries to retain them. We probe this
behaviour using a time-stepping algorithm. The time-averaged norm for
several $\mu$ is given in Fig.~\ref{fig:condil} and one can well
appreciate that the corresponding solution branch emerges from the
saddle-node bifurcation at $\mu_c$ indicating that it is actually a
Saddle Node Infinite PERiod (SNIPER) bifurcation. This is furthermore
corroborated by the square root dependence of the inverse time-period
(mean sliding speed) on $\mu-\mu_c$ that is given in the inset of
Fig.~\ref{fig:condil} \cite{Stro94,ThKn06}.
An example of a time series of snapshots for a stick-slip motion of a
single drop is given in Fig.~\ref{fig:snapsnipil}. Note that the times
at which the snapshots are taken are not equidistant.  It takes the
drop about 25000 time units to slowly stretch away from the defect
(snapshot 1 to 2). Then within 500 units it depins and slides to the
next defect (snapshot 2 to 5), where it needs another 25000 units to
reach an identical state as in snapshot 1 (snapshot 5 to 6). The
depinning itself resembles a pinch-off event at a water tap: the
bridge between drop and a 'reservoir' on the hydrophilic stripe
becomes thinner until it snaps. Once the main drop slides a small drop
remains behind on the defect.  All together for the chosen value of
$\mu$ the ratio of stick/stretch and slip phase is about $50:1$. The
ratio diverges when approaching the bifurcation.

Next we discuss the case of a hydrophobic
defect. Fig.~\ref{fig:condob} is the corresponding bifurcation
diagram. It shows as solid line stable solutions corresponding to
single drops blocked at their front by the line defect with
$\epsilon=0.3$. Dashed lines indicate unstable steady solutions. The
general behaviour resembles strongly the related 2d case and as well
the hydrophilic case. In particular, does the norm of the stable
solutions increase with increasing $\mu$ as the drop is increasingly
pushed against the defect and becomes therefore steeper. The drop
itself becomes more oval as its transverse width increases but the
streamwise one decreases.  An example of such a stable steady drop can
be seen in the left panel of Fig.~\ref{fig:statob}. The two other
panels represent the two unstable solutions that exist for identical
$\mu$ (crosses in Fig.~\ref{fig:condob}). Both unstable drops are
situated mainly upstream of the defect but have downstream protrusions
of different length and strength that reach the substrate beyond the
defect. The drop on the middle branch corresponds to a threshold
solution as in the hydrophilic case.

Time simulations indicate that depinning occurs again via a sniper
bifurcation, i.e., a branch of time-periodic solutions emerges from
the saddle-node at $\mu_c$.  However, the time-period does not diverge
as $(\mu-\mu_c)^{-1/2}$ but rather with the power $-1/4$ (inset of
Fig.~\ref{fig:condob}).
An example of a time series of snapshots for a stick-slip motion of a
depinned drop is given in Fig.~\ref{fig:snapsnipob}.  The drop needs
about 1600 time units to slowly let a 'protrusion' creep over the
defect (snapshot 1 to 2). Then within 400 units it depins and slides
to the next defect (snapshot 2 to 5), where it needs another 1200
units to reach the state as in snapshot 1 (snapshot 5 to 6). Then the
cycle starts again.  All together for the chosen value of $\mu$ the
ratio of stick and slip phase is about $7:1$. Once the drop is
depinned a small drop is retained behind the defect (snapshot 4).

Comparing the 2d and 3d cases we find that the depinning behaviour for
drops of equal height agrees qualitatively, but quantitatively there
is a small systematic difference.  In both cases we find depinning
transitions via a sniper bifurcation at a critical driving
$\mu_c$. However, in 3d $\mu_c$ is about 10-15\% larger than the one
in 2d. This is a result of the smaller mass per lateral length the 3d
drop has as compared to the ridge. Such an effect increases $\mu_c$ as
it implies a smaller ``effective 2d loading'' in the 3d
case. Actually, from the dependency on loading in 2d
(see~\cite{ThKn06}) one would expect an even larger difference. The
reason for the small increase may be the additional degree of freedom
that a true 3d drop has as compared to a translationally invariant
ridge.  It allows the drop to 'probe' the barrier locally by an
advancing protrusion (in the hydrophobic case) or by thinning its
backward bridge to the defect (in the hydrophilic case). It can
therefore use a pathway of morphological changes for depinning that a
2d drop is not able to use. Note that individual stations of this
pathway that can be seen in Figs.~\ref{fig:snapsnipil} and
\ref{fig:snapsnipob} do very much resemble the unstable steady
solutions presented in Figs.~\ref{fig:statil} and \ref{fig:statob},
respectively.  This indicates that the steady solutions that exist
below $\mu_c$ are still present in the phase space as 'ghost
solutions' \cite{Stro94} and can be seen in the course of the time
periodic motion beyond $\mu_c$.  When we have discussed that the
depinning behaviour for $\epsilon=\pm0.3$ in 2d and 3d is very similar
we have focused on the branch of pinned drops only. Note that the
connectivity of the unstable branches is not that similar. In this
respect the 3d case resembles the 2d case at a larger contrast
$\epsilon$.

Finally, we discuss the character of the single remaining steady state
solution for large $\mu$. Comparing 
bifurcation diagrams for large increasing $\mu$ (not shown) one notes
that in the 2d case the norm approaches zero and the solutions
resemble slightly modulated films. In contrast, in 3d the norm
approaches a finite value, i.e., there remains a non-trivial large
amplitude structure. The character of this structure can be
appreciated in Fig.~\ref{fig:rivulet}. Equally for hydrophilic as for
hydrophobic defects one finds a rivulet with drop-like transverse
cross sections and comparatively small variation in streamwise
direction.  Increasing $\mu$ the streamwise modulation becomes even
smaller and the thickness profiles in the transverse direction are
very close to steady 2d drops on horizontal substrates of
corresponding wettability. The rivulet is linearly unstable below a
large finite driving $\mu_r$. There it stabilizes via a Hopf
bifurcation that as well forms the end point of the branch of
time-periodic solutions.

%
%
%


%

\section{Conclusion}

We have studied depinning three-dimensional drops under lateral
driving for localized hydrophobic and hydrophilic line defects
employing on the one hand continuation techniques to obtain
steady-state solutions (pinned drops and rivulets) and their stability
and on the other hand a time-stepping algorithm to study the dynamics
of the stick-slip motion beyond depinning.
We have found that for the studied parameter range the depinning
behavior is qualitatively similar in 2d and 3d: Drops are pinned up to
a critical driving $\mu_c$ where they depin via a sniper
bifurcation. Quantitatively there exists a small systematic difference
-- the 3d $\mu_c$ is slightly larger than the 2d one.  Our
interpretation is that the difference results mainly from a lower
``effective 2d loading'' in the 3d case but is has as well to be taken
into account that the 3d drop has additional degree of freedom
enabling it to employ pathways of morphological changes for depinning
that a 2d drop is not able to access.

The sniper bifurcation is in the hydrophilic case characterized by a
square-root power law dependence of the inverse time scale of
depinning on the distance from threshold $\mu-\mu_c$. Beyond
$\mu_c$ the unsteady motion resembles the stick-slip motion observed
in experiment: The advance of the drop is extremely slow when it
'creeps away' from the hydrophilic region, and very fast once the
thread connecting the back of the drop to the defect has broken and
the drop slides to the next defect. The difference in time scales for
the stick and the slip phase diverges when approaching $\mu_c$. For a
hydrophilic defect, however, we have found a degenerate sniper
bifurcation as the power law has an exponent of about 1/4.
Re-considering the 2d case we found that there as well in the
hydrophobic case the power is about 1/3, i.e., it differs from the
expected 1/2 (cf.~\cite{ThKn06}). This may result from a degeneracy of
the problem that has, however, still to be identified.

Note that for hydrophobic defects of large strength depinning may
occur at very large driving via a Hopf instead of a sniper bifurcation
\cite{ThKn06b}. Then depinning is caused by the flow in the
wetting layer. For realistic forces the effect can not be observed for
partially wetting nano- or micro-drops on an incline or rotating
disc and we have not considered the parameter regime here.  Note,
however, that micro-drops of dielectric liquids generated by an
electric field in a capacitor can coexist with a thick wetting layer
of 100nm to 1$\mu$m stabilized by van der Waals interaction
\cite{Lin01,MPBT05}. In such a setting both depinning mechanisms
should be observable using gravity as the driving force (see appendix
of \cite{ThKn06}).




\acknowledgments
We acknowledge support by the EU [MRTN-CT-2004005728 PATTERNS] and the DFG
[SFB 486, B13].


\end{document}